\documentclass[amsmath,amssymb,11pt]{article}
\usepackage{myjheppub}
\pdfoutput=1
\usepackage{graphicx,epsfig}
\usepackage{float}
\usepackage{subfig}
\usepackage{subfloat}
\usepackage[utf8]{inputenc}
\usepackage{hyperref}
\usepackage{caption}
\usepackage{color}
\usepackage{calc}

\usepackage{soul}

\newcommand{\beq}{\begin{equation}}

\newcommand{\eeq}{\end{equation}}

\renewcommand{\=}{\; = \;}

\newsavebox\CBox
\newcommand\hcancel[2][0.5pt]{%
  \ifmmode\sbox\CBox{$#2$}\else\sbox\CBox{#2}\fi%
  \makebox[0pt][l]{\usebox\CBox}%
  \rule[0.75\ht\CBox-#1/2]{\wd\CBox}{#1}}

\newcommand{\ext}{\text{extr}}
\newcommand{\lp}{\ell_{R}}
\newcommand{\lm}{\ell_{L}}
\newcommand{\tllm}{\widetilde{\ell}_{L}}
\newcommand{\tllp}{\widetilde{\ell}_{R}}

\title{Gravitational index of the D1-D5-P black string}

\author{Silvia Georgescu,}
\emailAdd{silvia.georgescu@kcl.ac.uk}
\author{Sameer Murthy,}
\emailAdd{sameer.murthy@kcl.ac.uk}
\author{and Andrew Svesko}
\emailAdd{andrew.svesko@kcl.ac.uk}

\affiliation{
Department of Mathematics, King's College London, Strand, London, WC2R 2LS, UK}

\abstract{We present saddle-points of the Euclidean Gravitational Path Integral (GPI) corresponding to the supersymmetric
index of the D1-D5-P black string. 
These saddles are complex, supersymmetric, non-extremal solutions of 10-dimensional IIB supergravity theory 
with arbitrary inverse temperature $\beta$. 
The solutions carry fixed monopole charges $Q_1$, $Q_5$, $Q_n$, and angular momentum~$J_L$ equal to the extremal supersymmetric black string. 
Crucially, the solutions have fixed angular velocity $\Omega_{R} = - 2 \pi i /\beta$,
which is the chemical potential dual to~$J_R$.
This implements the insertion of~$(-1)^{F}$ in the GPI.
The Gibbons-Hawking on-shell action is temperature-independent
and agrees with the supersymmetric index of the extremal black string carrying the same monopole charges.
Upon taking a finite-temperature near-horizon decoupling limit, we obtain  
solutions of the form $S^3$ fibered over BTZ. 
Although these near-horizon solutions have finite holomorphic and anti-holomorphic modular parameters, 
their on-shell action reproduces the Cardy formula of the 
holomorphic elliptic genus of the dual D1-D5 SCFT$_2$.
}

\begin{document}

\maketitle

\section{Introduction}

The \textit{gravitational index} in supersymmetric theories of gravity is 
the gravitational path integral (GPI) \`a la Gibbons-Hawking~\cite{Gibbons:1976ue}, but with asymptotic boundary 
conditions appropriate to the calculation of a supersymmetric index~\cite{Cabo-Bizet:2018ehj,Iliesiu:2021are,Boruch:2023gfn} 
(see~\cite{Cassani:2025sim} for a review).
In situations in which the gravitational theory has a dual description in terms of a weakly-coupled 
microscopic theory, the gravitational index describes the direct strongly-coupled macroscopic analogue of 
the supersymmetric index of the microscopic dual.
In this paper, we present saddle-point solutions to the gravitational index 
corresponding to the D1-D5-P black string in string theory. 

\medskip

\noindent \textbf{Gravitational index and supersymmetric black holes} 

\noindent The notion of gravitational index helps to clarify some puzzling aspects of the explanation
of the entropy of supersymmetric black holes (BH) in string theory.
Recall that, in a class of compactifications of superstring theory, the supersymmetric index of the microscopic theory, 
calculated using a weakly-coupled ensemble of strings and branes~\cite{Sen:1995in,Strominger:1996sh},
contains a growth of states corresponding precisely to the entropy of the supersymmetric black hole in the 
dual gravitational theory. 
The first puzzling aspect of this agreement is that the index is a difference between the number of bosons and fermions,
while the entropy should correspond to a total number of states. This aspect per se can be explained 
by showing that all microstates are bosonic in the near-horizon of the BH~\cite{Sen:2009vz,Dabholkar:2010rm},
or, equivalently, to a vanishing theorem for the index (see~\cite{Murthy:2023mbc} for a discussion). 
However, there is a more direct puzzle: while the microscopic supersymmetric index can be defined at any 
temperature~$1/\beta$, and its value is independent of the temperature~\cite{Witten:1982df}, supersymmetric extremal black 
holes necessarily have zero temperature. So the two calculations do not even formally correspond to 
the same observable.  

The gravitational index resolves this puzzle as follows. 
The true saddle-points of the gravitational index are smooth non-extremal complex 
configurations\footnote{These saddles have been  
shown to be consistent with the
Konstevich-Segal-Witten allowability criterion~\cite{Kontsevich:2021dmb, Witten:2021nzp} for complex saddles to the 
GPI~\cite{BenettiGenolini:2025jwe, BenettiGenolini:2026raa, Krishna:2026rma}.} 
that preserve supersymmetry with asymptotic size~$\beta$ of the Euclidean time circle. 
Although the horizon values of the area and the moduli fields of the true saddle are~$\beta$-dependent, 
it is a non-trivial fact that the Gibbons-Hawking on-shell action of these saddles that defines the  gravitational 
free energy is independent of~$\beta$ and of the moduli (apart from a trivial factor coming from the 
Legendre transform). 
Upon taking the limit~$\beta \to \infty$ of the Legendre transform, one obtains the entropy of the  supersymmetric extremal black hole. 
The independence of the gravitational index from~$\beta$ and the moduli is the gravitational 
analogue of the mechanism of the Witten index in the microscopic theory.

\medskip

\noindent \textbf{Boundary conditions for the gravitational index} 

\noindent The asymptotic boundary conditions for the index are 
implemented in the gravitational path integral as follows~\cite{Cabo-Bizet:2018ehj, Iliesiu:2021are,Boruch:2023gfn}. 
Typically, the theories of supergravity with supersymmetric black hole solutions contain 
symmetry generators that do not commute with the supercharge that defines the index. 
These could be angular momentum or R-symmetry gauge fields of the supergravity, which we generically 
call R-symmetries of the black hole.\footnote{This nomenclature is partly motivated by the fact that these generators, 
whether spin or internal R-symmetry, indeed form part of the R-symmetry group in the near-horizon region of the black hole.} 
The idea is to turn on a holonomy, at asymptotic infinity, of the gauge field that carries such an 
R-charge.\footnote{The gauge field could be a spacetime gauge field, as in the case of R-symmetry 
transformations in supergravity, or it could be an off-diagonal component of the metric that couples to angular 
momentum, or a combination of the two~\cite{Cabo-Bizet:2018ehj}.}
This gives a certain potential to the R-charge. Since fermions are charged under the R-symmetry, 
the holonomy effectively implements supersymmetric boundary conditions.\footnote{One could restate this in 
geometric terms~\cite{Iliesiu:2021are}. Taking the R-symmetry to be angular momentum for the moment, it is clear that 
the holonomy is equivalent to a twist such that the contractible cycle is a combination of the thermal cycle and an 
angular direction. Then one demands that the fermions are periodic on the thermal circle (and hence supersymmetric), 
but are antiperiodic (and hence smooth) on the contractible cycle. In fact, in string theory, the R-symmetry gauge field 
of supergravity is typically realized as the spin in an internal circle of compactification, in which case there is a 
geometric picture in string theory.}

\smallskip

Another way to think about this is via the Gibbons-Hawking thermodynamic interpretation of the GPI. 
We use spacetime angular momentum~$J$ 
and its dual potential, i.e.,~the angular velocity~$\Omega$ to illustrate the point, but it applies more generally to any R-charge.
Recall that, according to the gravitational thermodynamics interpretation, 
the GPI is supposed to compute a thermal trace including coupling to external chemical 
potentials appropriate to the ensemble. Now, it is easy to see that going to the grand-canonical 
ensemble for angular momentum with the value~$\Omega = \pm 2 \pi i/\beta$ has the effect of inserting a $(-1)^{F}$ 
into the gravitational partition function, resulting in the supersymmetric index. We can summarize this as follows, 
\beq 
\text{Tr} \, \bigl( e^{-\beta H} e^{-\beta \Omega J}\dots\bigr)\Big{|}_{\Omega=\pm \frac{2\pi i}{\beta}}
\=\text{Tr}\big((-1)^{F}e^{-\beta H}\dots\big)\;,
\label{eq:gravtosusyind}\eeq 
where the ellipsis 
denote the effect of other charges carried by the configuration. 
The advantage of the above idea of implementing the index is that one has converted a fermionic problem 
to a bosonic problem of solving equations of supergravity with certain boundary conditions. 
Of course, one still needs to solve these equations and find a smooth solution. 
The holonomy of the smooth R-symmetry gauge field around a contractible cycle at infinity implies the 
presence of corresponding flux in the geometry. 
The index-saddles therefore contain R-charge different from the corresponding black hole solution.
In the simplest case, in four-dimensional~$\mathcal{N}=2$ ungauged supergravity, 
supersymmetric black holes do not rotate, while the corresponding index-saddles carry spin.

\medskip
\noindent \textbf{Index saddles.} 
The non-extremal index-saddle of the gravitational index regulates the IR singularity of the extremal 
infinite throat in a manifestly supersymmetric manner. 
Such index-saddles corresponding to supersymmetric black holes have been found in various theories of 
supergravity in asymptotic~$\mathbb{R}^d$ as well as in asymptotic AdS$_{d+1}$ 
space~\cite{Cabo-Bizet:2018ehj,Bobev:2020pjk,Boruch:2025qdq,Boruch:2025biv,Boruch:2025sie,Adhikari:2024zif,
Anupam:2023yns,Larsen:2026sav,Hegde:2024bmb,Chen:2024gmc, Iliesiu:2021are,
Cassani:2024kjn,BenettiGenolini:2023ucp,BenettiGenolini:2023rkq,Cassani:2021dwa,Cassani:2025iix}. 
Whenever there is a weakly coupled dual (given by a theory of strings 
and branes in~$\mathbb{R}^d$ or by a dual CFT$_d$ in the two situations, respectively), one sees an agreement 
of the microscopic and the gravitational index, confirming the formal predictions of supersymmetry. 
Similar index-saddles of other black objects like multi-center black holes, 
black strings, black lenses, black saturns have been studied 
in~\cite{Boruch:2025qdq,Boruch:2025biv,Cassani:2025iix,Boruch:2025sie,Bandyopadhyay:2025jbc,
Nanda:2026mbp,Dharanipragada:2026dji}.

\medskip

\noindent \textbf{The D1-D5-P black string.} In this paper, we focus on the  D1-D5-P black string in string theory, the first system in which the microscopic calculation of the black hole entropy was performed~\cite{Strominger:1996sh}. 
The theory we study can be regarded as the compactification of Type II string theory on~$M_4 = T^4$. The string solutions carry left- and right-moving angular momenta called~$J_L$, $J_R$. 
In order to obtain saddle-point solutions to the gravitational index, 
in accord with the discussion above, we turn on the angular velocity~$\Omega_R$ dual to~$J_R$ such that~$\beta \, \Omega_R = -2 \pi i$.\footnote{The sign in this equation is a choice, and throughout this paper we pick the negative sign.}

We present index saddles for the non-extremal black string in~$\mathbb{R}^6$. Upon compactification on one circle, these map to the five-dimensional non-extremal black string index saddles~\cite{Boruch:2025qdq}.
Further, we consider a decoupling limit in which we obtain non-extremal supersymmetric
solutions of the type $S^3$ fibered over BTZ, which are saddle-points of the elliptic genus of the boundary SCFT$_2$ of the D1-D5 system.

\medskip

\noindent \textbf{Plan of the article.} The rest of this article is organized as follows. In Section~\ref{sec:d1d5pBS} we review the rotating D1-D5-P black string solution, including its gravitational thermodynamics and evaluation of the on-shell Euclidean Type IIB supergravity action (computational details are relegated to Appendix~\ref{app:onshellactD1D5P}). 
We determine the gravitational index saddle for the D1-D5-P black string in Section~\ref{sec:indexd1d5p}. We further take a finite temperature decoupling limit, yielding non-extremal supersymmetric $S^3$ fibered over BTZ solutions, and show the agreement between the gravitational index and the superconformal index of the dual D1-D5 CFT.

\vspace{2mm} 

\noindent \textbf{Note:} As our article was in preparation we became aware of the related work \cite{Nanda:2026mbp}, 
with which Section~\ref{sec:indexd1d5p} has  overlap.

\section{The D1-D5-P black string}\label{sec:d1d5pBS}

In string theory black holes arise as bound states of intersecting D-branes wrapped on internal compact spaces.  The standard example is the 
three charge five-dimensional (5D) 
black hole solution to $\mathcal{N}=2$ supergravity with three $U(1)$ gauge fields, the Cvetic-Youm solution \cite{Cvetic:1996xz}. 
Special limits of the solution include the Breckenridge-Myers-Peet-Vafa (BMPV) solutions~\cite{Breckenridge:1996is,Breckenridge:1996sn} and the Strominger-Vafa black hole \cite{Strominger:1996sh}.

In \cite{Cvetic:1998xh} the generic Cvetic-Youm solution was interpreted as a 6D rotating black string, 
related to the 5D black hole via a dimensional reduction over the circle wrapped by the string~$S_{y}^{1}$ 
(the ``string circle" from now on).
The prototype system is the D1-D5-P 
black string, which can be viewed as a solution to 10D 
Type IIB supergravity on $\mathbb{R}^{1,4}\times S_y^{1}\times T^{4}$. 
From the 10D perspective, the construction involves $N_{5}$ D5-branes wrapped on the full compact 
space $S_{y}^{1}\times T^{4}$, $N_{1}$ D1-branes around the $S_{y}^{1}$-circle and uniformly smeared 
over the torus, along with $N_{n}$ units of momentum along the circle. 
In this section we review this black string solution and its gravitational thermodynamics. 

\subsection{The gravitational solution}

The non-extremal, rotating 10D metric in string frame is
\beq 
ds^{2}_{10} \= ds^{2}_{6}+\sqrt{\frac{H_{1}}{H_{5}}}(dx_{6}^{2}+dx_{7}^{2}+dx_{8}^{2}+dx_{9}^{2})\;,
\eeq
where the six-dimensional line element has the form of a black string \cite{Cvetic:1998xh,Jejjala:2005yu}, 
\beq 
\begin{split}
 ds^{2}_{6} \= &\frac{1}{\sqrt{H_{1}H_{5}}}\biggr\{-\left(1-\frac{r_{0}^{2} f_{D}}{r^{2}}\right)d\tilde{t}^{2}
 +d\tilde{y}^{2}+H_{1}H_{5}f^{-1}_{D}\frac{r^{4}}{(r^{2}+\ell_{1}^{2})(r^{2}+\ell_{2}^{2})-r_{0}^{2}r^{2}}dr^{2}\\
 &-\frac{2r_{0}^{2} f_{D}}{r^{2}}\cosh\alpha_{1}\cosh\alpha_{5}(\ell_{2}\cos^{2}\theta d\psi+\ell_{1}\sin^{2}\theta d\phi)d\tilde{t}\\
 &-\frac{2r_{0}^{2} f_{D}}{r^{2}}\sinh\alpha_{1}\sinh\alpha_{5}(\ell_{1}\cos^{2}\theta d\psi+\ell_{2}\sin^{2}\theta d\phi)d\tilde{y}\\
 &+\left[\left(1+\frac{\ell^{2}_{2}}{r^{2}}\right)H_{1}H_{5}r^{2}+(\ell_{1}^{2}-\ell_{2}^{2})
 \cos^{2}\theta\left(\frac{r_{0}^{2} f_{D}}{r^{2}}\right)^{2}\sinh^{2}\alpha_{1}\sinh^{2}\alpha_{5}\right]\cos^{2}\theta d\psi^{2}\\
 &+\left[\left(1+\frac{\ell^{2}_{1}}{r^{2}}\right)H_{1}H_{5}r^{2}+(\ell_{2}^{2}-\ell_{1}^{2})
 \sin^{2}\theta\left(\frac{r_{0}^{2} f_{D}}{r^{2}}\right)^{2}\sinh^{2}\alpha_{1}\sinh^{2}\alpha_{5}\right]\sin^{2}\theta d\phi^{2}\\
 &+\frac{r_{0}^{2}f_{D}}{r^{2}}\left(\ell_{2}\cos^{2}\theta d\psi+\ell_{1}\sin^{2}\theta d\phi\right)^{2}
 +H_{1}H_{5}r^{2}f_{D}^{-1}d\theta^{2}\biggr\}\;,
\end{split}
\label{eq:CYmet}\eeq
with metric functions
\beq 
\begin{split}
&H_{1} \= 1+\frac{r_{0}^{2}f_{D}\sinh^{2}\alpha_{1}}{r^{2}}\;,
\qquad H_{5} \= 1+\frac{r_{0}^{2} f_{D}\sinh^{2}\alpha_{5}}{r^{2}}\;,\\
&f_{D}^{-1} \=1+\frac{\ell_{1}^{2}\cos^{2}\theta+\ell_{2}^{2}\sin^{2}\theta}{r^{2}}\;,
\end{split}
\eeq
and boosted coordinates
\beq 
d\tilde{t}\=\cosh\alpha_{n} dt-\sinh\alpha_{n}dy\;,\qquad 
d\tilde{y}\=\cosh\alpha_{n} dy-\sinh\alpha_{n} dt\;.
\eeq
Here $\theta\in (0,\pi/2)$, $\{\psi,\phi\}\in (0,2\pi)$, coordinates $x_{6},\dots,x_{9}$ characterize the torus $T^{4}$, with $x_{a}\sim x_{a}+2\pi R_{a}$ ($a=6,\dots,9$), and the coordinate $y$ of the string circle $S^{1}_y$ obeys $y\sim y+2\pi R$.

The $C_{(2)}$ gauge field supporting the solution is, up to a constant gauge transformation,
\beq 
\begin{split}
 C_{(2)}\= &-\frac{r_{0}^{2} f_{D}\cos^{2}\theta}{H_{1}r^{2}}\left[(\ell_{1}c_{1}s_{5}c_{n}-\ell_{2}s_{1}c_{5}s_{n})dt
 	+(\ell_{2}s_{1}c_{5}c_{n}-\ell_{1}c_{1}s_{5}s_{n})dy\right]\wedge d\psi\\
 &-\frac{r_{0}^{2} f_{D}\sin^{2}\theta}{H_{1}r^{2}}\left[(\ell_{2}c_{1}s_{5}c_{n}-\ell_{1}s_{1}c_{5}s_{n})dt
 	+(\ell_{1}s_{1}c_{5}c_{n}-\ell_{2}c_{1}s_{5}s_{n})dy\right]\wedge d\phi\\
 &-\frac{r_{0}^{2} f_{D} s_{1}c_{1}}{H_{1}r^{2}}dt\wedge dy-\frac{r_{0}^{2} f_{D}s_{5}c_{5}}{H_{1}r^{2}}(r^{2}
 	+\ell_{1}^{2}+r_{0}^{2}s_{1}^{2})\cos^{2}\theta d\psi\wedge d\phi\;,
\end{split}
\label{eq:C2gffull}\eeq
where we use the notation 
\beq 
s_{i}\=\sinh\alpha_{i}\;,\qquad c_{i}\=\cosh\alpha_{i}\;,\qquad i\=1,5,n \;. 
\eeq
The field strength $F_{(3)}=dC_{(2)}$ is computed to be 
 \begin{align}\label{eq:fieldstrengthgen}
  F_{(3)} &\= 2r_{0}^{2}r\cos^{2}\theta\left(\frac{f_{D}}{H_{1}r^{2}}\right)^{2}dr\wedge 
  	\left[(\ell_{1}c_{1}s_{5}c_{n}-\ell_{2}s_{1}c_{5}s_{n})dt+(\ell_{2}s_{1}c_{5}c_{n}-\ell_{1}c_{1}s_{5}s_{n})dy\right]\wedge d\psi\nonumber\\
  &+2r_{0}^{2}\sin\theta\cos\theta\left(\frac{f_{D}}{H_{1}r^{2}}\right)^{2}(\ell_{2}^{2}+r^{2}+r_{0}^{2}s_{1}^{2})d\theta\wedge \nonumber\\
&\qquad \qquad \qquad \qquad \qquad \qquad  \left[(\ell_{1}c_{1}s_{5}c_{n}-\ell_{2}s_{1}c_{5}s_{n})dt+
(\ell_{2}s_{1}c_{5}c_{n}-\ell_{1}c_{1}s_{5}s_{n})dy\right]\wedge d\psi\nonumber\\
  &+2r_{0}^{2}r\sin^{2}\theta\left(\frac{f_{D}}{H_{1}r^{2}}\right)^{2}dr\wedge\left[(\ell_{2}c_{1}s_{5}c_{n}-\ell_{1}s_{1}c_{5}s_{n})dt+(\ell_{1}s_{1}c_{5}c_{n}-\ell_{2}c_{1}s_{5}s_{n})dy\right]\wedge d\phi\nonumber\\
  &-2r_{0}^{2}\sin\theta\cos\theta\left(\frac{f_{D}}{H_{1}r^{2}}\right)^{2}(\ell_{1}^{2}+r^{2}+r_{0}^{2}s_{1}^{2})d\theta \wedge \nonumber\\
&\qquad \qquad \qquad \qquad \qquad \qquad  \left[(\ell_{2}c_{1}s_{5}c_{n}-\ell_{1}s_{1}c_{5}s_{n})dt
	+(\ell_{1}s_{1}c_{5}c_{n}-\ell_{2}c_{1}s_{5}s_{n})dy\right]\wedge d\phi\nonumber\\
  &+2r_{0}^{2}rs_{1}c_{1}\left(\frac{f_{D}}{H_{1}r^{2}}\right)^{2}dr\wedge dt\wedge dy-2r_{0}^{2} s_{1}c_{1}\sin\theta\cos\theta(\ell_{1}^{2}-\ell_{2}^{2})\left(\frac{f_{D}}{H_{1}r^{2}}\right)^{2}d\theta \wedge dt\wedge dy\nonumber\\
  &+2r_{0}^{2} c_{5}s_{5}(\ell_{2}^{2}+r^{2}+r_{0}^{2}s_{1}^{2})(\ell_{1}^{2}+r^{2}+r_{0}^{2}s_{1}^{2})\sin\theta\cos\theta\left(\frac{f_{D}}{H_{1}r^{2}}\right)^{2}d\theta\wedge d\psi\wedge d\phi\nonumber\\
  &+2r_{0}^{2}rs_{5}c_{5}(\ell_{1}^{2}-\ell_{2}^{2})\left(\frac{f_{D}}{H_{1}r^{2}}\right)^{2}\sin^{2}\theta\cos^{2}\theta dr\wedge d\psi \wedge d\phi\;.
 \end{align}
The dilaton $\Phi$ is given by
\beq 
e^{2\Phi} \= g_{s}^{2}\frac{H_{1}}{H_{5}}\;,
\eeq
where~$g_s$ is the ten-dimensional string coupling at infinity.
The 5D, 6D, and 10D Newton constants, $G_{5},G_{6},G_{10}$, are related via 
\beq 
(2\pi)^{5}VR \, G_{5} \= (2\pi)^{4}VG_{6} \= G_{10} \= 8\pi^{6}g_{s}^{2}\alpha'^{4}\;,
\label{eq:Newtconsts}\eeq
where the volume of the~$T^4$ is parameterized by~$V=R_{6}R_{7}R_{8}R_{9}$. 
In what follows we will set $G_{5}=\pi/4$. After a reduction on the $T^{4}$, the above configuration is a solution to the 6D $\mathcal{N}=4$ maximal supergravity.

\vspace{2mm}

\noindent \textbf{Conserved charges.} The above solution is characterized by six parameters 
$r_{0},\ell_{1},\ell_{2},\alpha_{1},\alpha_{5},\alpha_{n}$, which are related to the six conserved charges as follows.  The ADM mass is given by 
\beq \label{ADMmass}
M\= r_{0}^{2}\,\Bigl(\,\frac{3}{2}+ s_1^2+s_5^2+s_n^2\Bigr) \;. 
\eeq
The angular momenta associated to rotations along $\phi$ and $\psi$ are 
\beq
   J_{\phi}\= r_{0}^{2}\bigl(\ell_{1}c_{1}c_{5}c_{n}-\ell_{2}s_{1}s_{5}s_{n}\bigr)\;,\qquad 
   J_{\psi}\= r_{0}^{2}\bigl(\ell_{2}c_{1}c_{5}c_{n}-\ell_{1}s_{1}s_{5}s_{n}\bigr)\;,
\eeq
and the corresponding left and right-moving combinations are given by
\beq \label{JLR}
    J_{L}\=\frac{1}{2}(J_{\phi}-J_{\psi}) \=\frac{r_{0}^{2}}{2} \, \lm \, \bigl(c_{1}c_{5}c_{n}+s_{1}s_{5}s_{n} \bigr)\;,\qquad 
 J_{R}\=\frac{1}{2}(J_{\phi}+J_{\psi}) \=\frac{r_{0}^{2}}{2} \, \lp \, (c_{1}c_{5}c_{n}-s_{1}s_{5}s_{n} \bigr)\;,
\eeq
where we have introduced the combinations 
\beq
\lm \= \ell_1 - \ell_2 \;, \qquad 
\lp \= \ell_1 + \ell_2 \;.
\label{eq:lrlL}\eeq
The above solution has a symmetry under the simultaneous interchange $\ell_1\leftrightarrow \ell_2$~, $\phi\leftrightarrow\psi$ and $\theta\leftrightarrow \frac{\pi}{2}-\theta$. It is also invariant under the simultaneous transformations $\ell_{1}\rightarrow -\ell_{1}$, $\phi\rightarrow -\phi$, $\ell_{2}\rightarrow -\ell_{2}$ and $\psi\rightarrow -\psi$. In the Lorentzian theory, without loss of generality, we set $\ell_1\geq \ell_2\geq 0$, such that $\lp \ge \lm$. 
This choice will affect the expressions for the horizon radii and the thermodynamic quantities below.

Finally, the electric and magnetic charges associated to the 3-form and momentum along the string circle are 
\beq \label{chargesQ}
Q_{1}\=r_0^2 \, s_1 \, c_1\;,\qquad 
Q_{5}\= r_0^2 \, s_5 \, c_5\;,\qquad 
Q_{n}\=r_0^2 \, s_n \, c_n\;,\eeq
In string theory, these charges are quantized as\footnote{Here we used our convention $G_5=\pi/4$, which sets $\frac{g_s^2\alpha'^4}{V R}=1$.} $Q_1=\frac{g_{s}\alpha'^{3}}{V} N_1$, $Q_5=g_{s}\alpha' N_5$, 
and $Q_n= \frac{N_n}{R}$ where $N_1$, $N_5$, $N_n$ denotes the number of 
D1 branes, D5 branes, and units of momentum supporting the solution, respectively.
We take~$N_i$ to be positive throughout the paper, which 
translates to $\alpha_{i}>0$ for $i=1,5,n$.\footnote{Changing the signs of the charges can lead to a 
different extremal solution see, e.g.,~\cite{Goldstein:2008fq}.}

\vspace{2mm}

\noindent \textbf{Horizons and entropy.} 
The inner and outer horizons, determined by
$g(r)\equiv \frac{\sqrt{H_{1}H_{5}}}{f_{D}} r^4 g^{rr} = (r^{2}+\ell_{1}^{2})(r^{2}+\ell_{2}^{2})-r_{0}^{2}r^{2}=0$, for inverse metric component $g^{rr}$,  are  
\beq r_{\pm} 
\= \frac{1}{2}\left(\sqrt{r_{0}^{2}-\lm ^{\,2}}\pm \sqrt{r_{0}^{2}-\lp^{\,2}}\right)\;,
\label{eq:horrad}\eeq
where $r_{+}\geq r_{-}\geq0$, a consequence of our choice below (\ref{eq:lrlL}). 
Rearranging, we obtain
\beq 
r_{0}^{2} \= \frac{1}{r_{+}^{2}}(r_{+}^{2}+\ell_{1}^{2})(r_{+}^{2}+\ell_{2}^{2})\;.
\label{eq:defmpar}\eeq
To avoid a conical defect, naked singularity, or (smooth) soliton, one further needs 
\beq \label{eq:regularity}
r_{0}^{2} \; \geq \; \lp^{\, 2}\;.
\eeq
The Bekenstein-Hawking area-entropy is given by 
\beq 
S_{\text{BH}} \=
\pi r_{0}^{2}\Bigl( \bigl(c_{1}c_{5}c_{n}+s_{1}s_{5}s_{n} \bigr)\sqrt{r_{0}^{2}-\lm^{2}}
+\bigl(c_{1}c_{5}c_{n}-s_{1}s_{5}s_{n}\bigr)\sqrt{r_{0}^{2}-\lp^{2}} \;\Bigr)\;.
\label{eq:nonextentrop}\eeq

\subsection{Thermodynamics}
 
We now briefly recall the thermodynamics of this black string solution  following
Appendix~A of~\cite{Dias:2007dj}. 
Firstly, note that the horizon is generated by the Killing vector 
\beq 
\xi^{\mu} \= \partial_{t}+\Omega_{\phi}\partial_{\phi}+\Omega_{\psi}\partial_{\psi}\;,
\eeq
which becomes null at $r=r_{+}$. Here,  
\beq 
\Omega_{\phi}\=\frac{\ell_{1}r_{+}^{2}}{(r_{+}^{2}+\ell_{1}^{2})(r_{+}^{2}c_{1}c_{5}c_{n}+\ell_{1}\ell_{2}s_{1}s_{5}s_{n})}\;,\qquad \Omega_{\psi}\=\frac{\ell_{2}r_{+}^{2}}{(r_{+}^{2}+\ell_{2}^{2})(r_{+}^{2}c_{1}c_{5}c_{n}+\ell_{1}\ell_{2}s_{1}s_{5}s_{n})}\;
\eeq
are the angular velocities at the horizon. These quantities can be considered as the chemical potentials 
coupling to the angular momenta in the black hole thermodynamics. 
The combinations that couple to the left and right angular momenta~\eqref{JLR} are, 
\beq 
\begin{split}
\Omega_{L}& \;\equiv \; \Omega_{\phi}-\Omega_{\psi} \= 
\frac{\lm\, r_{+}^{2}(r_{+}^{2}-\ell_{1}\ell_{2})}{(r_{+}^{2}+\ell_{1}^{2})(r_{+}^{2}
+\ell_{2}^{2})(r_{+}^{2}c_{1}c_{5}c_{n}+\ell_{1}\ell_{2}s_{1}s_{5}s_{n})} \;,
\end{split}
\eeq
\beq
\begin{split}
\Omega_{R}& \; \equiv \; \Omega_{\phi}+\Omega_{\psi}
\=\frac{\lp\, r_{+}^{2}(r_{+}^{2}+\ell_{1}\ell_{2})}{(r_{+}^{2}+\ell_{1}^{2})(r_{+}^{2}
+\ell_{2}^{2})(r_{+}^{2}c_{1}c_{5}c_{n}+\ell_{1}\ell_{2}s_{1}s_{5}s_{n})} \;,
\end{split}
\eeq
where we use (\ref{eq:horrad}) and (\ref{eq:defmpar}). 

The potentials at the horizon are
\beq \Phi_{i}\=\frac{r_{+}^{2}\frac{s_i}{c_i}c_{1}c_{5}c_{n}+\ell_{1}\ell_{2}\frac{c_i}{s_i}s_{1}s_{5}s_{n}}{r_{+}^{2}c_{1}c_{5}c_{n}+\ell_{1}\ell_{2}s_{1}s_{5}s_{n}}\;, \qquad i=1,5,n\;.
\eeq
Here~$\Phi_1$, $\Phi_5$ are the electric potentials at the horizon associated with the gauge charges $Q_{1}$, $Q_5$, respectively, 
and~$\Phi_n$ is the velocity along the string circle, which is the potential associated with the momentum~$Q_n$ along~$S^1_y$.

Finally, the inverse Hawking temperature $\beta=T^{-1}$ is given by 
\begin{align}\label{eq:inversetemp}
    \beta&\=\frac{1}{2}\bigg(\underbrace{\frac{2\pi r_0^2 (c_1 c_5 c_n-s_1 s_5 s_n)}{\sqrt{r_0^2-\lm^2}}}_{\beta_L}+\underbrace{\frac{2\pi r_0^2 (c_1 c_5 c_n+s_1 s_5 s_n)}{\sqrt{r_0^2-\lp^2}}}_{\beta_R}\bigg) \;.
    \end{align}
It will be useful to express the chemical potentials in terms of $\beta$, 
\begin{align}\label{potentialsgeneral}
\Omega_L&\=\frac{2\pi}{\beta} \frac{
\lm}{\sqrt{r_0^2-\lm^2}},\hspace{1cm}
\Omega_R\=\frac{2\pi}{\beta} \frac{\lp}{\sqrt{r_0^2-\lp^2}} \; ,\\
\Phi_i&\=\frac{\pi r_0^2}{\beta}\bigg( \frac{\frac{s_i}{c_i}c_1 c_5 c_n - \frac{c_i}{s_i}s_1 s_5 s_n}{\sqrt{r_0^2-\lm^2}}+ \frac{\frac{s_i}{c_i}c_1 c_5 c_n + \frac{c_i}{s_i}s_1 s_5 s_n}{\sqrt{r_0^2-\lp^2}} \bigg) \; , \qquad i=1,5,n\;.
\end{align}

It is straightforward to verify that the above thermodynamic variables obey the 
first law of thermodynamics, i.e.,  
\beq 
dM\=TdS_{\text{BH}}+\sum_{i=1,5,n}\Phi_{i} \, dQ_{i}+\Omega_{L}\, dJ_{L}+\Omega_{R}\, dJ_{R}\;.
\eeq
The grand canonical free-energy is equal to 
\beq \label{freeenergy}
\mathcal{F}(\Phi_{i},\Omega_{L,R},T)\=M-TS_{\text{BH}}-\sum_{i=1,5,n}\Phi_{i} \, Q_{i}-(\Omega_{L}J_{L}+\Omega_{R}J_{R})\;,
\eeq
and the dual variables are related to each other as 
\beq
\left(\frac{\partial \mathcal{F}}{\partial \Phi_i}\right)_{T,\Omega_{L,R}} \= -Q_i\;, \quad \left(\frac{\partial \mathcal{F}}{\partial \Omega_{L,R}}\right)_{T,\Phi_{i},} \= -J_{L,R}\;,\quad \left(\frac{\partial \mathcal{F}}{\partial T}\right)_{\Phi_{i},\Omega_{L,R}} \= -S_{\text{BH}}\;.
\eeq
Lastly, we observe the following simple relation, 
\beq \mathcal{F}+\Phi_{i}Q_{i}=\frac{r_{0}^{2}}{2}\cosh 2\alpha_{i}\;,\quad i=1,5,n\;,\label{eq:useidfreeen}\eeq
which holds with and without rotation and will be useful in the following.

\subsection{On-shell action}\label{Section:onshellaction}
Here we derive the free energy (\ref{freeenergy}) via a saddle-point approximation of the gravitational partition function $\mathcal{Z}$ using the Gibbons-Hawking prescription, where
\begin{align}
    \beta \mathcal{F}&=-\log\mathcal{Z}=I_{E}^{\mathrm{on-shell}}\;,
\end{align}
for the on-shell Euclidean action $I^{\text{on-shell}}_{E}$.

The theory of interest is Type IIB supergravity, characterized by the Lorentzian action, in Einstein frame,
\begin{align}\label{typeiibaction}
I^{\text{bulk}}&=\frac{1}{2\kappa_{10}^2}\int d^{10}x \sqrt{-g}\bigg(R-\frac{1}{2}\nabla_{\mu}\Phi\nabla^{\mu}\Phi-\frac{1}{12}e^{\Phi}F_{(3)}^{\mu\nu\rho}F_{\mu\nu\rho}^{(3)}\bigg) \;,
\end{align}
where $\kappa_{10}^2=8\pi G_{10}$. The equations of motion are
\begin{align}
&R_{\mu\nu}-\frac{1}{2}g_{\mu\nu}R-\frac{1}{2}\left(\nabla_{\mu}\Phi\nabla_{\nu}\Phi-\frac{1}{2}g_{\mu\nu}\nabla_{\mu}\Phi\nabla^{\mu}\Phi\right)-\frac{1}{4}e^{\Phi}\left(F^{(3)}_{\mu\alpha\beta}F_{\nu}^{(3)\alpha\beta}-\frac{1}{6}g_{\mu\nu}F_{(3)}^{\alpha\beta\gamma}F^{(3)}_{\alpha\beta\gamma}\right)=0\;,\nonumber\\
&\Box\Phi-\frac{1}{12}e^{\Phi}F_{(3)}^{\mu\nu\rho}F_{\mu\nu\rho}^{(3)}=0\;,\\
&\nabla_{\mu}(e^{\Phi}F^{\mu\alpha\beta})=0\nonumber\;.
\end{align}
From the equations of motion we have
\begin{align}
    R&=\frac{1}{2}(\Box\Phi + \nabla_{\mu}\Phi\nabla^{\mu}\Phi)\;.
\end{align}
Applying the equations of motion and Wick rotating the time coordinate $t=-it_{E}$, for real time $t_{E}$, the Euclidean bulk action, on-shell, is
\begin{align}
I^{\text{bulk}}_{E}&=\frac{1}{4\kappa_{10}^{2}}\int d^{10}x\sqrt{g}\Box\Phi=\frac{1}{4\kappa_{10}^{2}}\int d^{9}y\sqrt{h}n^{\mu}\nabla_{\mu}\Phi\biggr|_{r=r_{+}}^{\infty}\;.
\label{eq:bulkactonshell}\end{align}
To obtain the second equality we performed an integration by parts along the radial direction, 
such that the remaining contribution evaluates to a term on the spacetime boundary, 
including contributions from a constant-radial hypersurface at the horizon and spatial infinity. Here $n^{\mu}$ is the outward pointing unit normal at the boundary with induced metric $h_{\mu\nu}$. Evaluating this bulk term for the (Euclideanized) D1-D5-P black string gives
\begin{align}\label{bulkterm}
     I^{\text{bulk}}_{E}&=\frac{\beta r_0^2}{4}\left(\sinh^2\alpha_5-\sinh^2\alpha_1\right)\;.
\end{align}

To have a well-posed variational principle, the bulk action \eqref{typeiibaction} needs to be supplemented by the Gibbons-Hawking-York (GHY) boundary term. In Euclidean signature,
\begin{align}
    I^{\text{GHY}}_{E}&=-\frac{1}{\kappa^2_{10}}\int d^{9}y\sqrt{h } (K-K_0)\biggr|_{r=\infty}\;,
\label{eq:GHY}\end{align}
where the spacetime boundary has trace of extrinsic curvature $K$. 
We use the method of background subtraction to regularize divergences the GHY term has at spatial infinity. 
The final regularized result for the GHY term is (see Appendix \ref{app:onshellactD1D5P} for details)
\begin{align}
 I^{\text{GHY}}_{E}&= \frac{\beta r_0^2}{8}(\cosh 2\alpha_1+3\cosh 2\alpha_5)\;.
\label{eq:GHYonshell}\end{align}
Combined with the bulk contribution, the total on-shell Euclidean action is
\begin{align}\label{resultonshellaction}
    I^{\text{on-shell}}_E&= I^{\text{bulk}}_{E}+ I^{\text{GHY}}_{E}=\frac{\beta r_0^2}{2}\,\cosh 2\alpha_5\;.
\end{align}
Note that the result does not depend on the rotation parameters $\ell_{L,R}$. Let us now compare with the free energy \eqref{freeenergy}.

Recall that the GHY boundary term in the action makes the variational principle well-posed assuming the induced metric obeys Dirichlet boundary conditions. Since $\Phi_n,\Omega_{L,R}$ are given by off-diagonal elements in the metric, the total action (\ref{resultonshellaction}) automatically computes the free energy in a thermal ensemble with $\Phi_n,\Omega_{L,R}$ fixed. Additionally, we take the electric potential $\Phi_{1}$ to be fixed. 
As such we did not include a Maxwell-like boundary term for the $C_{(2)}$ form, i.e.,  we did not fix the number of D1 branes. 
Note, however, the number of D5 branes enters the magnetic charge $Q_{5}$ which is automatically fixed, even without a boundary term \cite{Hawking:1995ap}. 
Our computation above thus corresponds to a thermal ensemble with $\{\Omega_{L,R},\Phi_{1,n},Q_{5}\}$ fixed, and the on-shell action computes the free energy
\begin{align}
    I_{E}^{\text{on-shell}}&=\beta(\mathcal{F}+\Phi_5 Q_5)\nonumber\\
    &=\beta M - S_{\text{BH}} - \beta\Omega_L J_L- \beta\Omega_R J_R - \beta\Phi_1 Q_1- \beta\Phi_n Q_n\;.
\end{align}
Indeed, comparing the identity \eqref{eq:useidfreeen} with on-shell action \eqref{resultonshellaction}, we find perfect agreement.\footnote{In principle, one can work in the democratic formalism, which introduces the dual potential $C_{(6)}$ and  treats $Q_1$ and $Q_5$ in the same way, in order to reproduce the full grand canonical free energy. We will not pursue this approach here.}

\section{Index of the D1-D5-P black string}\label{sec:indexd1d5p}

In this section we determine the gravitational index saddle for the D1-D5-P black string.
To do so, we use the general black string solution described in the previous section, and impose conditions of supersymmetry appropriate to the index, i.e.,~$\beta\,\Omega_{R}=-2\pi i$ in the Euclidean ensemble. 
This procedure leads to a one-parameter family of supersymmetric solutions labelled by the non-extremality parameter~$\beta$. 
In particular, this leads us to a certain scaling regime in the parameter space of the solution. 

We begin by describing the extremal and supersymmetric solution in Lorentzian signature.
We then go back to the general non-extremal solution and impose the index condition.
This leads us to a particular scaling regime in the parameters. 
Finally, we present the finite temperature decoupling limit of the index saddle, 
in which we obtain a complex Euclidean BTZ black hole with~$S^3$ fibered over it. 
We end with a brief discussion of how this index saddle contributes to the gravitational index dual to the elliptic genus of the boundary~SCFT$_2$. 

\subsection{The extremal supersymmetric solution \label{sec:extremalsusy}}

\noindent \textbf{Extremality.} 
The extremal solution is defined by the limit 
when the inner and outer horizons coincide,  $r_{+}=r_{-}\equiv r_{\ext}$. 
Equivalently, we have $g(r)=0$ and $g'(r)=0$ in terms of the function~$g$ defined above~\eqref{eq:horrad}. 
We summarize this as follows,
\beq
r_{0}^{2}\=\lp^{2} \;,  \qquad r_{\ext}\=\sqrt{\ell_{1}\ell_{2}}\;.
\qquad\qquad
\label{eq:extcond}\eeq
Recall that we have made a choice described below (\ref{eq:lrlL}) according to which  
the inverse right-moving temperature $\beta_{R}$ in (\ref{eq:inversetemp}) diverges to positive infinity (and hence $\beta \to \infty$) in the extremal solution.

In this limit, only five of the parameters are independent, and we can choose them to 
be~$\alpha_{1,5,n},\lp,\lm$. 
The thermodynamic variables of the black string can then be expressed as, 
\beq 
\begin{split}
&T_{\ext}\=0\;, \qquad S^{\ext}_{\text{BH}}\=2\pi \,\lp^{2} \,(c_{1}c_{5}c_{n}+s_{1}s_{5}s_{n})\, r_{\ext}\;,\\
&\Omega^{\ext}_{L}\=0\;, \qquad \Omega^{\ext}_{R}\=\frac{2}{\lp(c_{1}c_{5}c_{n}+s_{1}s_{5}s_{n})}\;,\qquad\Phi^{\ext}_{i}\=\frac{\frac{s_i}{c_i}c_{1}c_{5}c_{n}+\frac{c_i}{s_i}s_{1}s_{5}s_{n}}{c_{1}c_{5}c_{n}+s_{1}s_{5}s_{n}}\;.
\end{split}
\eeq 
Equivalently, in the limit, only five of the  conserved charges are independent. In particular, 
we express~$J_R$ as, 
\begin{align}
J^{\ext}_{R}\=\frac{\lp^{3}}{4}(c_{1}c_{5}c_{n}-s_{1}s_{5}s_{n})\;.
\end{align}
Further, it is straightforward to verify
\beq 
\begin{split}
 S^{\ext}_{\text{BH}}\=2\pi \sqrt{Q_{1}Q_{5}Q_{n}-J_{L}^{2}+J_{R}^{2}}\;.
\end{split}
\label{eq:SBHext}\eeq

\vspace{2mm}

\noindent \textbf{Supersymmetry.} 
The Type IIB supersymmetry algebra allows for BPS configurations preserving 4 supercharges 
when the following condition is obeyed, 
\beq \label{BPScondition}
M \= Q_{1}+Q_{5}+Q_{n}\;.
\eeq
Note that rotation does not explicitly appear in the above BPS condition. 
Upon expressing the charges in terms of the parameters of the black string solution using~\eqref{ADMmass} and \eqref{chargesQ},
we see that the condition~\eqref{BPScondition} can be satisfied with finite, non-zero charges only if
$r_0=0$ and $\alpha_i\rightarrow +\infty$ 
with the product~$r_0 \, e^{\alpha_i}$ held fixed.
We take such a limit of the parameters below.

Upon further imposing the Lorentzian regularity condition~\eqref{eq:regularity}, we find 
that~$\lp=0$, 
which brings us to the extremal solution above with~$J_R=0$. 
In this limit, the other thermodynamic quantities take the following values,
\begin{align} \label{eq:extsusyvals}
    \beta\=\infty\;,\qquad\Omega_{L,R}\=0\;,
    \qquad\Phi_i\=1 \;,\quad i\=1,5,n \;.
\end{align}
The extremal supersymmetric solution that we thus obtain has four independent parameters and, equivalently, 
four independent conserved charges~$Q_i$, $i=1,5,n$, and~$J_L$. 

To summarize, supersymmetry combined with regularity in the Lorentzian theory leads us to the extremal black string.

\subsection{Index condition}

Now, instead of imposing regularity in the Lorentzian solution, we perform a Wick rotation of the time coordinate, 
and impose conditions on the chemical potentials in the non-extremal solution to obtain the index saddle. 
As explained in the 
introduction, we impose 
\beq 
\beta \,\Omega_{R} \; = \; -2\pi i\;,
\label{eq:SUSYcond}\eeq
which has the effect of inserting a $(-1)^{F}$ into the gravitational partition function, resulting in 
the supersymmetric index, 
cf.~Eqn.~(\ref{eq:gravtosusyind}). 
For the (Lorentzian) non-extremal Cvetic-Youm geometry, we have, using (\ref{potentialsgeneral}),
\beq 
  \beta \, \Omega_{R}
  \= 2\pi \frac{\lp}{\sqrt{r_{0}^{2}-\lp^{2}}}\;.
\eeq
Upon imposing (\ref{eq:SUSYcond}), we find $r_{0}=0$, with the choice $\sqrt{-\lp^2}=i \lp$. 
As anticipated, this implies complexification of certain parameters, which are not allowed in the Lorentzian solution, but are saddle-points to the ``Euclidean" GPI.

From the mass~\eqref{ADMmass} and gauge charges~\eqref{chargesQ}, 
we see that naively setting $r_{0}=0$ leads to vanishing conserved charges $M$ and $Q_i$. 
A way to remedy this is to simultaneously take 
\beq 
 r_{0}\to 0\;,\qquad \alpha_{i}\to \infty\;,\qquad 
\frac{r_{0} \, e^{\alpha_{i}}}{2} \;= \;\; \text{fixed}\;,\qquad i\=1,5,n \;,
\label{eq:susylimv1}\eeq
with $G_{10},V$ and $R$ fixed. Applying this limit to \eqref{ADMmass} and \eqref{chargesQ} shows that  the mass~$M$ and the gauge charges~$Q_i$ are nonzero and satisfy the BPS condition~\eqref{BPScondition}, as expected. 
This limit is the same as the one we impose in the extremal condition above, but now we have reached it as a consequence of supersymmetry alone for non-zero finite charges.

Notice, however, the angular momentum~$J_{L}$ \eqref{JLR} diverges as~$r_0^{-1}$ in the limit~(\ref{eq:susylimv1}). To keep $J_{L}$ finite, we introduce
\begin{align}\label{leftrescaling}
   \tllm \; \equiv \; \frac{\lm}{r_0}\;,
\end{align}
and require $\tllm$ to be finite in the limit (\ref{eq:susylimv1}). Thus,
\begin{align}
    J_L \; \rightarrow \; \tllm \sqrt{Q_1 Q_5 Q_n} \;.
\end{align}

On the other hand, for finite $\lp$, the right angular momentum~$J_R$ vanishes.
Indeed, in the limit~(\ref{eq:susylimv1}), one has $c_1 c_5 c_n-s_1 s_5 s_n\sim r_0^{-1}$, and~\eqref{JLR} leads to 
\begin{align}\label{vanishingJR}
    J_R \; \rightarrow \; 0\;.
\end{align}
For finite $\lp$, we have $\beta_R \rightarrow\infty$, and hence $\beta\rightarrow\infty$.
The entropy also reduces to the extremal and supersymmetric result (\ref{eq:SBHext})
\begin{align}
    S_{\text{BH}} \; \rightarrow \; 2\pi \sqrt{Q_{1}Q_{5}Q_{n}-J_{L}^{2}}\;.
\end{align}

The solution that we have reached in the scaling limit is essentially identical to the original extremal black string~\cite{Cvetic:1996xz}.
The one difference is that we still have the additional free parameter~$\lp$ in our solutions. 
In the extremal solution~\cite{Cvetic:1996xz}, both~$\lm$ and~$\lp$ are taken to zero with  $\lm/r_{0}$ and $\lp/r_{0}$ held fixed in the limit. 
But, in fact, all the thermodynamic expressions including the charges and entropy are independent of~$\lp$ in this limit, as long as~$\lp$ is finite. 

As we explain below, we use this extra ``degree of freedom" in the solutions to obtain a non-extremal index saddle. 
To do so, we need to take~$\lp \to \infty$ in such a way that $\beta$ is finite and $J_R$ is non-zero. 
In fact, this limit  coincides exactly with the rescaling used in~\cite{Anupam:2023yns}.

\subsection{Non-extremal index saddles}

From the expression of the right- and left-moving temperatures (\ref{eq:inversetemp}), in the limit~\eqref{eq:susylimv1}, with rescaling \eqref{leftrescaling}, we see $\beta_L$ is finite while
\begin{align}
    \beta_R \; \rightarrow \; \frac{2\pi \sqrt{Q_1 Q_5 Q_n}}{i \, r_0 \, \lp} \;.
\end{align}
To keep~$\beta_R$ finite in the supersymmetric solution, we further rescale~$\lp \to \infty$ such that  
\begin{align}
    \tllp \; \equiv \; r_0 \, \lp
\label{eq:scalinglp}\end{align}
is fixed. This is precisely the scaling required for the index saddle, which is both supersymmetric 
and defined for arbitrary $\beta$. 
Moreover, we have 
\begin{align}
    c_1 c_5 c_n - s_1 s_5 s_n \= \frac{1}{2 r_0}\frac{Q_S}{\sqrt{Q_1 Q_5 Q_n}} \;,
\end{align}
where, in the notation of~\cite{Anupam:2023yns}, 
\begin{align}
    Q_S\=Q_1 Q_5 +Q_1 Q_n+Q_5 Q_n \;.
\end{align}
It is easy to see the scaling (\ref{eq:scalinglp}) then yields a nonvanishing right-moving angular momentum~\cite{Anupam:2023yns}
\begin{align}\label{RMangularmomentum}
    J_R\=\frac{\tllp}{4} \frac{Q_S}{\sqrt{Q_1 Q_5 Q_n}} \;.
\end{align}
This allows us to trade $\tllp$ for the conserved charges. 

The final expression for the full inverse temperature is
\begin{align}\label{inversetemp}
    \beta \= \frac{\pi Q_S}{2}\Biggl( \,
\frac{1}{\sqrt{Q_1Q_5Q_n-J_L^2}}
-\frac{i}{J_R} \, \Biggr)\;,
\end{align}
and the expression for the entropy is
\begin{align}
    S_{\text{BH}} \= 2\pi \,\Bigl( \sqrt{Q_1 Q_5 Q_n-J_L^2} \; + \; 2\pi i J_R \Bigr) \;.
\end{align}
Requiring the entropy to be real forces $J_R$, and therefore $\tllp$, to be purely imaginary. With this choice $\beta$ is also real while $\Omega_R$ 
must be purely imaginary such that~\eqref{eq:SUSYcond} is  satisfied. 

The limiting values of the rest of the thermodynamic quantities are given by
\begin{align}
    \Omega_L&\=\frac{4 i J_L J_R}
{Q_S\left(\sqrt{Q_1Q_5Q_n-J_L^2}+iJ_R\right)}\;,\qquad \Omega_R\=\frac{4 J_R\sqrt{Q_1Q_5Q_n-J_L^2}}
{Q_S\left(\sqrt{Q_1Q_5Q_n-J_L^2}+iJ_R\right)} \;,\nonumber\\
    \Phi_i&\=1-\frac{1}{Q_i}\frac{2 i J_R Q_1 Q_5 Q_n}
{Q_S \left(\sqrt{Q_1Q_5Q_n-J_L^2}+iJ_R\right)} \;.
\end{align}
The supersymmetric solution has five independent parameters, which we take to be the four conserved charges~$J_L$, $Q_i$, $i=1,5,n$, plus a fifth parameter which we can take to be~$J_R$ or~$\beta$. 
In the limit $\beta\rightarrow\infty$, one recovers the extremal supersymmetric solution 
discussed in Section~\ref{sec:extremalsusy} and the parameters take the values~\eqref{eq:extsusyvals}.

Finally, the thermodynamic grand canonical free-energy \eqref{freeenergy} is  
\begin{align}
   \beta \, \mathcal{F}\=\frac{\pi Q_1 Q_5 Q_n}
{\sqrt{Q_1 Q_5 Q_n-J_L^2}} \;.
\label{eq:freenind}\end{align}
We can check this is indeed reproduced by the Euclidean on-shell action computed in Section~\ref{Section:onshellaction}, 
after performing a Legendre transform to the grand-canonical ensemble and taking the supersymmetric limit.

The grand-canonical partition function and the corresponding on-shell action~$I$ can be written 
in  terms of the potentials as follows, 
\begin{align}\label{onshellactionpotentials}
      I \; \equiv \; \beta \, \mathcal{F} \= -\frac{4 \varphi_{1} \varphi_{5} \varphi_{n}}
{\omega_L^2-\omega_R^2} \;,
\end{align}
where we introduced
\begin{align}\label{susypotentials}
    \varphi_i \;\equiv \;\beta(\Phi_i -1)=-\frac{1}{Q_i}\frac{\pi Q_1 Q_5 Q_n}
{\sqrt{Q_1Q_5Q_n-J_L^2}}\;,\qquad
\omega_L \; \equiv \; \beta \, \Omega_L \;, \qquad 
\omega_R \; \equiv \; \beta \, \Omega_R \;,
\end{align}
such that $\omega_{R}=-2\pi i$.

\vspace{2mm}

\noindent \textbf{Index solution.} Let us now consider the limit (\ref{eq:susylimv1}) on the solution. 
Firstly we notice from~\eqref{eq:horrad} that the horizon radius $r_+$ diverges as  
\begin{align}\label{divrplus}
    r_+ \; \sim \; \frac{i}{2}\frac{\tllp}{r_0} \;.
\end{align}
In fact, as originally noted by \cite{Anupam:2023yns}, the metric (\ref{eq:CYmet}) is not finite in the limit (\ref{eq:susylimv1}). 
A finite metric can be obtained by changing the radial coordinate~\cite{Anupam:2023yns,Nanda:2026mbp} 
\begin{align}
    \rho^2&\;\equiv \;r^2 -\frac{r_0^2- (\lp^2 + \lm^2)/2}{2} \;.
\end{align}
At leading order in the $r_0\rightarrow 0$ limit, with the quantities previously specified fixed, we obtain
\begin{align}
    \rho^2&\=r^2+\frac{{\tllp}^{\,2}}{4 r_0^2}+\text{O}(1)\;,
\end{align}
such that $\rho(r_+)=\text{O}(1)$, where the divergence \eqref{divrplus} is precisely canceled.

The resulting metric, upon Wick rotating $t=-it_{E}$, is\footnote{The solution matches that in Appendix A.2 of \cite{Nanda:2026mbp} with $y_{\text{here}}=-y_{\text{there}}$, $J_{L,R}^{\text{here}}=2J_{L,R}^{\text{there}}$ and $f_{1,5}=\tilde{H}_{1,5}$.} 
\begin{align}
    &ds^2=\frac{1}{\sqrt{f_1 f_5}}
\bigg[
(f-Q_n)\,dt_{E}^2
+(f+Q_n)\,dy^2
+2iQ_n\,dt_{E}\,dy
+\frac{f_1 f_5\,d\rho^2}
{\rho^2-\dfrac{4J_R^2\left(J_L^2-Q_1Q_5Q_n\right)}
{Q_S^2\rho^2}}
\nonumber\\
&+f_1f_5\,d\theta^2+f_1\,ds^2_{T^4}-2i(J_L-J_R)\cos^2\theta\,d\psi\,dt_{E}
+2i(J_L+J_R)\sin^2\theta\,d\phi\,dt_{E}
\nonumber\\
&+2\left(J_R-J_L-\frac{2J_RQ_1Q_5}{Q_S}\right)\cos^2\theta\,dy\,d\psi
+2\left(J_L+J_R-\frac{2J_RQ_1Q_5}{Q_S}\right)\sin^2\theta\,dy\,d\phi\nonumber\\
&+\frac{2J_R^2Q_1Q_5Q_n\sin^2 2\theta}{Q_S^2}d\psi d\phi+\left(
f_1 f_5+\frac{4J_R\cos^2\theta\left(J_RQ_1Q_5Q_n-J_LQ_S(f+Q_1+Q_5)\right)}
{Q_S^2}
\right)
\cos^2\theta\,d\psi^2\nonumber\\
&+\left(
f_1 f_5
+
\frac{4J_R\sin^2\theta\left(J_RQ_1Q_5Q_n+J_LQ_S(f+Q_1+Q_5)\right)}
{Q_S^2}
\right)
\sin^2\theta\,d\phi^2\bigg]\;,
\label{eq:indexsol}\end{align}
where $f= \rho^2 + \frac{2 J_L J_R \cos2\theta}{Q_S}$ and $f_{1,5}=f + Q_{1,5}$.\footnote{Recall that we have so far suppressed an overall factor of $\frac{g_s^2\alpha'^4}{V R}$ in the expression for~$Q_i$ given in~\eqref{chargesQ}.
In other words, $Q_i$ in the metric above stands for~$Q_i/(\frac{g_s^2\alpha'^4}{V R})= r_0^2 s_i c_i$ in the supersymmetric limit. 
Reinstating the factors of $R$ will soon be important when we take the decoupling limit.} 

The 2-form potential is
\begin{align}
 &C_{(2)}=\frac{iQ_1}{f_1}dt_{E}\wedge dy-\frac{Q_5\left(Q_1+\rho^2+2J_LJ_R\right)\cos^2\theta}{f_1}d\psi\wedge d\phi\nonumber\\
    &+i\left(J_L+J_R-\dfrac{2J_RQ_1Q_n}{Q_S}\right)\frac{\cos^2\theta}{f_1} dt_{E}\wedge d\psi+\left(J_L-J_R+\dfrac{2J_RQ_5Q_n}{Q_S}\right)\frac{\cos^2\theta}{f_1} dy\wedge d\psi \nonumber\\
    &-i\left(J_L-J_R+\dfrac{2Q_1Q_nJ_R}{Q_S}\right)\frac{\sin^2\theta}{f_1} dt_{E}\wedge d\phi-\left(J_L+J_R-\dfrac{2Q_5Q_nJ_R}{Q_S}\right)\frac{\sin^2\theta}{f_1}dy\wedge d\phi\;,
\end{align}
and the dilaton is given by
\begin{align}
    e^{2\Phi}&=g_s^2 \frac{f_1}{f_5}\;.
\end{align}
Note that, although $J_R$ appears democratically with the other charges in the expressions above,
it can be thought of as a regulator of the extremal solution,
and it is distinct from the monopole charges of the extremal solution. 
In particular, $J_R$ is a function of $\beta$ and $J_L,Q_{1,5,n}$, according to~\eqref{inversetemp}.

\subsection{Decoupling at finite temperature} 

In the above subsection, we obtained the smooth, complex solution that is supersymmetric at finite temperature~$1/\beta$. 
In the limit~$\beta \to \infty$ we obtain the extremal black string solution with a near-horizon~AdS$_3$ region.

Now we take a  ``finite temperature'' decoupling limit of the index solution~(\ref{eq:indexsol}), following~\cite{Boruch:2025qdq}, in which we keep the temperature in the near-horizon region fixed. 
(The same scaling was also performed in~\cite{Nanda:2026mbp}). 
This allows us to directly relate the gravitational index to the microscopic index of the D1-D5 CFT. 
In order to formulate this limit, we introduce 
\begin{align}
    \ell_3&\equiv(Q_1 Q_5)^{1/4},
\end{align}
which becomes the radius of the AdS$_3$ in the near-horizon region. 
The decoupling limit corresponds to taking the radius of the string circle as well 
the radius of the Euclidean time circle to be large (in units of~$\ell_3$), 
while keeping their ratio fixed, i.e., 
\begin{align}
    \frac{R}{\ell_3}\rightarrow\infty\;,\qquad \frac{\beta}{\ell_3}\rightarrow\infty\;,\qquad \frac{\beta}{R}\quad \text{fixed}\;.
\end{align}
The charges $Q_{1,5}$ are kept fixed in the limit, 
while $Q_n\rightarrow 0$ as a consequence of $R\rightarrow\infty$.

It is convenient to introduce the dimensionless parameter~$\Lambda$. The decoupling limit corresponds to scaling 
\begin{align} \label{eq:decouplinglim}
   J_L \;\to\; \frac{J_L}{\Lambda},
\qquad
J_R \;\to\; \frac{J_R}{\Lambda},
\qquad
Q_n \;\to\; \frac{Q_n}{\Lambda^2},
\qquad
\rho \;\to\; \frac{\rho}{\Lambda},
\qquad
t_E \;\to\; \Lambda t_E,
\qquad
y \;\to\; \Lambda y
\end{align}
and taking the limit~$\Lambda\rightarrow\infty$. 
As explained in~\cite{Boruch:2025qdq} in the context of 5d black strings, 
the last two scalings are the coordinate versions of $\beta,R\rightarrow\infty$, since we prefer to keep the periodicities fixed. 
The scalings of the angular momenta follow from the dependence on $R$ (see \cite{Nanda:2026mbp} for more details), 
while the scaling of $\rho$ implements the near-horizon limit.

Applying the decoupling limit (\ref{eq:decouplinglim}) to the metric (\ref{eq:indexsol}) gives
    \begin{align}
    ds^2\= & \frac{\rho^2}{\ell_3^2}\,dt_E^2
+\frac{\rho^2}{\ell_3^2}\,dy^2
+\frac{\ell_3^{10}\rho^2}
{\ell_3^8\rho^4-4J_L^2J_R^2+4J_R^2\ell_3^4Q_n}\,d\rho^2 +\frac{\ell_3^4Q_n-J_L^2}{\ell_3^6}(i dt_E+dy)^2
\nonumber\\
&-\frac{J_R^2}{\ell_3^6}(i dt_E-dy)^2+\ell_3^2\Bigg[
d\theta^2
+\cos^2\theta
\left(
d\psi-\frac{i(J_L-J_R)dt_E+(J_L+J_R)dy}{\ell_3^4}
\right)^2\nonumber\\
&+\sin^2\theta
\left(
d\phi+\frac{(J_L-J_R)dy+i(J_L+J_R)dt_E}{\ell_3^4}
\right)^2
\Bigg]\;.
\label{eq:indexsaddledec1}\end{align}
In the $\rho\rightarrow\infty$ limit, the metric is locally AdS$_3\times S^3$. 

It is instructive to express the near-horizon metric in terms of the supersymmetric potentials~\eqref{susypotentials}, 
which map to the parameters in the supersymmetric index of the boundary theory on the boundary torus, i.e.,~the elliptic genus. 
To this end, note that in the decoupling limit~(\ref{eq:decouplinglim}) the charges $Q_{n}$ and $J_{L,R}$ may be expressed as
\begin{align}\label{chargesdec}
 Q_n\=\frac{\ell_3^4\left(\omega_L^2-\omega_R^2\right)}{4\varphi_n^2}\;,\qquad J_L\=\frac{\ell_3^4\omega_L}{2|\varphi_n|}\;,
\qquad
J_R\=\frac{\,\ell_3^4 \omega_R}{2(2\beta+\varphi_n)}\=-\frac{i\pi \,\ell_3^4}{2\beta+\varphi_n}\;.
\end{align}
To see this, apply the limit~(\ref{eq:decouplinglim}) 
to the potentials~(\ref{susypotentials}) 
such that $\varphi_{n}=-\frac{\pi}{\sqrt{\ell_{3}^{4}Q_{n}-J_{L}^{2}}}$ (after absorbing a factor of $\Lambda$ into $\varphi_{n}$) and $\frac{\omega_{L}}{\omega_{R}}=\frac{iJ_{L}}{\sqrt{\ell_{3}^{4}Q_{n}-J_{L}^{2}}}$. It is then straightforward to solve for the charges~$Q_{n},J_{L,R}$ in terms of the supersymmetric potentials.

We note that $\beta,\varphi_n,\omega_L$ are real, so that $Q_{n}$ and $J_{L}$ are real, while $J_{R}$ is complex. 
It is useful to write the left- and right-inverse temperatures as
\begin{align}\label{leftrighttemp}
\beta_L&\=|\varphi_n|,\qquad\beta_R\=2\beta+\varphi_n\;.
\end{align}
We consider $2\beta>|\varphi_n|>0$, as we are ultimately interested in taking $\beta\rightarrow\infty$.\footnote{In fact, we consider later in the computation $\beta>|\varphi_n|$, for the same reason.}

Using the charges (\ref{chargesdec}), the decoupled index saddle (\ref{eq:indexsaddledec1}) is\footnote{The solution matches Eqn. (3.33) from \cite{Larsen:2026sav}, with the identifications $ z_{\text{there}}=y_{\text{here}},\; \theta_{\text{there}}=2\theta_{\text{here}},\; \phi_{\text{there}}=(\psi-\phi)_{\text{here}},\; \psi_{\text{there}}=(\psi+\phi)_{\text{here}}$ and radial coordinate \eqref{newradialcoord}. The parameters are related by $\omega_L=-2\beta\,\tilde{\Phi}_R,\;\omega_R=2\beta\,\tilde{\Phi}_L,$ and $\varphi_n=\Omega'$.}
\begin{align}\label{eq:indexsadddecv2}
    ds^2&\=ds_3^2 + \ell_3^2\big(d\theta^2 + \cos^2\theta d\tilde{\psi}^2+\sin^2\theta d\tilde{\phi}^2\big),
\end{align}
where 
\begin{align}
\hspace{-1.5cm}   ds_3^2\= & \frac{\rho^2}{\ell_3^2}\,dt_E^2
+\frac{\rho^2}{\ell_3^2}\,dy^2
+\frac{\ell_3^2\pi^2}{(2\beta+\varphi_n)^2}(-i dt_E+dy)^2
\nonumber\\
& +\frac{\ell_3^2\pi^2}{\varphi_n^2}(i dt_E+dy)^2+
\frac{
\ell_3^2\rho^2(2\beta+\varphi_n)^2\varphi_n^2
}{\rho^4(2\beta+\varphi_n)^2\varphi_n^2-4\ell_3^8\pi^4
}\,d\rho^2\;,
\end{align}
and we have introduced 
\beq 
\begin{split} 
&d\tilde{\psi} \; \equiv \; 
d\psi
-\frac{\omega_R}{2(2\beta+\varphi_n)}(-i dt_E+dy)
+\frac{\omega_L}{2\varphi_n}(i dt_E+dy)\;,\\
&d\tilde{\phi}\; \equiv \; d\phi
-\frac{\omega_R}{2(2\beta+\varphi_n)}(-i dt_E+dy)
-\frac{\omega_L}{2\varphi_n}(i dt_E+dy)\;,
\end{split}
\eeq
for the $S^{3}$ fiber.

We can rewrite this saddle\footnote{The same decoupling limit can be taken for the $C_{(2)}$ potential and dilaton, but we do not present the details here.} in the canonical thermal AdS$_3$/BTZ form as follows. 
We change the radial coordinate to  
\begin{align}\label{newradialcoord}
    \tilde{\rho}&\=\frac{\ell_3}{2i}
\arccos\!\left(
\frac{\rho^2 \, |\varphi_n| \, (2\beta+\varphi_n)}
{2\, \ell_3^4 \, \pi^2}
\right) \;,
\end{align}
where we chose the integration constant in the definition of $\tilde{\rho}$ such that the horizon 
\begin{align}
    \rho_+&\=\frac{\sqrt{2}\pi\,\ell_3^2}
{\sqrt{|\varphi_n|}\,\sqrt{2\beta+\varphi_n}}\quad  \mapsto\quad \tilde{\rho}_+=0\;.
\end{align}
Consequently, the three-dimensional piece becomes 
\begin{align}\label{3dbtz}
ds^2_3 ={}& d\tilde{\rho}^{\,2}
+\cosh^2\!\left(\frac{\tilde{\rho}}{\ell_3}\right)d\eta^2+\sinh^2\!\left(\frac{\tilde{\rho}}{\ell_3}\right)d\xi^2\;,
\end{align}
where we have defined
\begin{align}
    \xi^2&\equiv\ell_3^2\pi^2\left(
\frac{t_E-i\,y}{\varphi_n}
-\frac{t_E+i\,y}{2\beta+\varphi_n}
\right)^2,\qquad\eta^2\equiv\ell_3^2\pi^2\left(
\frac{i\,t_E+y}{\varphi_n}+\frac{i\,t_E-y}{2\beta+\varphi_n}
\right)^2.
\label{eq:xietacoord}\end{align}

\vspace{4mm}

\noindent \textbf{Periodicity conditions and microscopic index.} The periodicity conditions of the decoupling limit of the index saddle were analyzed in detail in \cite{Larsen:2026sav}. 
Here we focus only on the three-dimensional metric. From the geometry \eqref{3dbtz}, we infer that the contractible circle corresponds to
\begin{align}
    (\xi,\eta)\sim(\xi+ 2\pi \ell_3, \eta)\;,
\end{align}
and is fixed by regularity around $\tilde{\rho}=0$. In terms of coordinates $(t_E,y)$\footnote{Although we focus on the three-dimensional part of the metric, the identifications required for the smoothness of the solution involve also shifts in the $S^3$ coordinates. For completeness and to avoid confusion, we therefore write the full identification.} the contractible circle is 
\begin{align}
    (t_E,y,\psi,\phi)\sim \big(t_E+\beta,y- i(\beta+\varphi_n),\psi-\pi,\phi-\pi\big)\;.
\end{align}
On the other hand, the non-contractible BTZ circle is
\begin{align}
    (t_E,y)\sim(t_E,y+2\pi R)\;.
\label{eq:noncontid}\end{align}
Therefore, the modular parameter $\tau$ characterizing the boundary torus of the asymptotic AdS$_{3}$ geometry, defined as the ratio of the non-contractible to the contractible circles of the boundary torus, is\footnote{To find the non-contractible circle, it suffices to substitute identification (\ref{eq:noncontid}) into coordinates (\ref{eq:xietacoord}), such that the periodicities of $\xi$ and $\eta$ are, respectively, $\Delta\xi=-\frac{4\pi^{2}i\ell_{3}R(\beta+\varphi_{n})}{\varphi_{n}(\varphi_{n}+2\beta)}$ and $\Delta\eta=-\frac{4\pi^{2}\ell_{3}R\beta}{\varphi_{n}(\varphi_{n}+2\beta)}$, where recall $2\beta+\varphi_{n}>0$ and $\varphi_{n}<0$. It follows that the modular parameter is $\tau=(\Delta\xi+i\Delta\eta)/(2\pi\ell_{3})$. }
\begin{align}
    \tau&=\frac{2\pi R}{i \varphi_n}\;,\qquad \bar{\tau}=\frac{2\pi R}{i(2\beta+ \varphi_n)}\;.
\label{eq:modparam}\end{align}
This agrees with the five-dimensional analysis (see Eqn.~(5.36) in~\cite{Boruch:2025qdq}) upon identifying $\varphi_n=\delta^0/2\pi$. Note that, using~\eqref{leftrighttemp}, $\tau=-\frac{2\pi R}{i \beta_L}$ and $\bar{\tau}=\frac{2\pi R}{i \beta_R}$.

Finally, we can connect to the Cardy formula as follows. In the decoupling limit, where we substitute the charges~(\ref{chargesdec}) into the free energy~(\ref{eq:freenind}), the on-shell action~\eqref{onshellactionpotentials} becomes
\begin{align}
     I \= \beta \mathcal{F}&\=\frac{\ell_3^4(\omega_L^2-
    \omega_R^2)}{4|\varphi_n|}\;,
\end{align}
which, in terms of charge $Q_n=N_n/R$, using~\eqref{chargesdec}, and modular parameter~(\ref{eq:modparam}), is
\begin{align}
      I&\=\frac{2\pi i N_n}{\tau}\;.
\end{align}
This agrees precisely with the Cardy formula for the growth of states of the elliptic genus of the boundary SCFT$_2$.\footnote{In particular, it corresponds to the so-called Type IIB Cardy limit, see the discussion in~\cite{Dabholkar:2010rm,Castro:2008ys}. 
This is consistent with the fact that we are studying the D1-D5 system in Type IIB theory.} 
We have reached this formula from a complex gravitational field configuration that is smooth and depends on~$\tau$ as well as~$\bar{\tau}$. 
But the saddle-point action is holomorphic in~$\tau$ as  consistent with the elliptic genus. 

\vspace{0.4cm}

Given the success in obtaining the index saddles to a variety of black holes and black strings, 
one may wonder if similar constructions exist for higher-dimensional branes. 
Another closely related brane solution to the one discussed in this article is its S-dual, namely the NS5-F1-P system. It is well known that the NS5-F1-P system admits an ``intermediate'' decoupling limit, in which one zooms into the near-horizon region of the NS5-branes, but not that of the F1-strings. The resulting three-dimensional geometry is not asymptotically AdS$_3$, but rather asymptotically flat with a linear dilaton. For the rotating NS5-F1-P solutions analogous to \eqref{eq:CYmet}, this intermediate decoupling limit was considered in \cite{Martinec:2018nco}. It would be interesting to find the corresponding index saddles and compute their on-shell action, which might provide insight into a notion of supersymmetric index for the dual non-gravitational theory.

\noindent\section*{Acknowledgments}

We are grateful to Weam Abou Hamdan, Roberto Emparan, Prakasam Shanmugapriya, and Amitabh Virmani for insightful feedback.
We thank Jan Boruch, Luca Iliesiu, and Dawid Maskalaniec for discussions on topics related to this paper. 
We thank all organisers and participants of the \href{https://www.ggi.infn.it/showevent.pl?id=543}{Pathways to Quantum Black Holes workshop} at the Galileo Galilei Institute for Theoretical Physics for stimulating discussions. S.G. is funded by a Royal Society Newton International Fellowship NIF/R1/241888. 
A.S. is funded by the Royal Society under the grant “Concrete Calculables in Quantum de Sitter” and  further supported by the STFC consolidated grant ST/X000753/1.
S.M.~acknowledges the support of the STFC grants ST/T000759/1,  ST/X000753/1 during the course of this work.

\appendix

\section{On-shell Euclidean action of D1-D5-P black string}\label{app:onshellactD1D5P}

Here we provide additional details on evaluating the on-shell action of the D1-D5-P black string as summarized in Section \ref{sec:d1d5pBS}.  

\vspace{2mm}

\noindent \textbf{Bulk contribution.} On-shell, the bulk contribution evaluates to a boundary term (\ref{eq:bulkactonshell})
\begin{align}
I^{\text{bulk}}_{E}&=\frac{1}{4\kappa_{10}^2}\int d^{10}x \sqrt{g}\Box\Phi
=\frac{1}{4\kappa_{10}^2}\int d^{9}y\sqrt{g}g^{rr}\partial_{r}\Phi\biggr|^{r=\infty}_{r=r_{+}}\;.
\end{align}
It is easy to see the bulk contribution vanishes at the horizon, such that the only non-trivial term arises from the quantity evaluated at spatial infinity. The integral over the Euclidean time circle $t_{E}$ produces the factor of $\beta$, while integrating over $S^{1}_{y}\times T^{4}$ produces the factor which changes $G_{10}$ to $G_{5}$ via the relation between Newton's constants (\ref{eq:Newtconsts}). The integration on the $S^3$ coordinates yields $V_{S^3}=2\pi^2$, and subsequently using our convention $G_{5}=\pi/4$ yields (\ref{bulkterm}).  

\vspace{2mm}

\noindent \textbf{Gibbons-Hawking-York contribution.} Consider the GHY boundary term (\ref{eq:GHY}) in the Euclideanized D1-D5-P background, neglecting, for now, the $K_{0}$ contribution. To proceed, first note that for this particular solution
\begin{align}
    K&=\nabla_{\mu}n^{\mu}=\frac{1}{\sqrt{g}}\partial_r( \sqrt{g} n^r)=\partial_r n^r +\frac{n^r}{2g}\partial_r g=\partial_r \bigg(\frac{1}{\sqrt{g_{rr}}}\bigg)+\frac{1}{2g\sqrt{g_{rr}}} \partial_r g\;.
\end{align}
Thus, we need only evaluate the determinant of the bulk metric and $g_{rr}$ metric component. Performing an expansion at a cutoff surface $r=r_{\text{M}}$ near the asymptotic boundary at spatial infinity, we have
\begin{align}
    K&= \frac{3}{r_{\text{M}}}
-\frac{
12\left(\ell_1^2+\ell_2^2-r_0^2\right)
+20\left(\ell_1^2-\ell_2^2\right)\cos 2\theta
+5r_0^2\left(\cosh 2\alpha_1+3\cosh 2\alpha_5 \right)}{
16r_{\text{M}}^3}+O(r_M^{-5})
\end{align}
The determinant of the induced metric at the boundary, meanwhile, has the expansion
\begin{align}
    \sqrt{h}&=r_{\text{M}}^3\sin\theta\cos\theta
+\frac{r_\text{M}}{16}\sin\theta\cos\theta
\biggr(
12\left(\ell_1^2+\ell_2^2-r_0^2\right)
\nonumber\\
&+4\left(\ell_1^2-\ell_2^2\right)\cos 2\theta+r_0^2\left(\cosh 2\alpha_1+3\cosh 2\alpha_5\right)
\biggr) +O(r_M^{-1})
\end{align}
Evidently, the GHY term diverges at spatial infinity, in the limit $r_{\text{M}}\to\infty$. 

To regulate this infrared divergence, we use the method of background subtraction. Specifically, we choose our reference background to be the (Euclidean) D1-D5-P solution with $\alpha_{1,5,n}=0$ and $r_0=0$ (having vanish gauge charges and ADM mass), however, we keep the rotation non-zero, $\ell_{1,2}\neq0$, i.e., 
\begin{align}
    ds^2_0&=dt_E^2 + dy^2 + \frac{
\tilde{r}^{\,2}\left(\tilde{r}^{\,2}
+\ell_1^2\cos^2\theta
+\ell_2^2\sin^2\theta\right)
}{
\left(\ell_1^2+\tilde{r}^{\,2}\right)
\left(\ell_2^2+\tilde{r}^{\,2}\right)
}d\tilde{r}^2 +(\tilde{r}^2+ \ell_1^2\cos^2\theta+\ell_2^2\sin^2\theta) d\theta^2 \nonumber\\
 &+(\tilde{r}^2+\ell_2^2)\cos^2\theta d\psi^2 +(\tilde{r}^2+\ell_1^2)\sin^2\theta d\phi^2 +dx_6^2+dx_7^2+dx_8^2+dx_9^2\;.
\end{align}
Further, we need the reference background to agree with the D1-D5-P solution along the cutoff surface at $r=r_{\text{M}}$ for large $r_{\text{M}}$.  To this end, we relate the radial coordinate $\tilde{r}$ to $r$ by
\begin{align}
    \tilde{r}^2&=r^2+\frac{1}{4}r_0^2\sinh^2\alpha_1
+\frac{3}{4}r_0^2\sinh^2\alpha_5\;.
\end{align}
With this change of radial coordinate, the trace of the extrinsic curvature of the subtracted background is 
\begin{align}
    K_0&=\frac{3}{r_\text{M}}
-\frac{4\left[3\left(\ell_1^2+\ell_2^2-r_0^2\right)
+5\left(\ell_1^2-\ell_2^2\right)\cos 2\theta
\right]+3r_0^2\cosh 2\alpha_1+9r_0^2\cosh 2\alpha_5}{16 r_{\text{M}}^3}+O(r_M^{-5})\,
\end{align}
in the large $r_{\text{M}}$ expansion. Consequently, 
\begin{align}
    K-K_0&=-\frac{r_0^2\left(\cosh 2\alpha_1+3\cosh 2\alpha_5\right)}{8r_M^3}+O(r_M^{-5})\;,
\end{align}
from which we obtain the finite on-shell result (\ref{eq:GHYonshell}).

\bibliography{refsindex}

\end{document}